# A putative, computationally stable structure of homotrimeric BP180/collagen XVII

Congzhou M Sha, MD, PhD[1,*]
[1]Penn Medicine Doylestown Hospital, 595 W State St, Doylestown, PA 18901
[*]Corresponding author. consha@sas.upenn.edu

Keywords: membrane-bound protein, bullous pemphigoid, BPAG2, Boltz-2, molecular dynamics

**Abstract (249 words)**

Background

BP180, also known as collagen XVII and BPAG2 (bullous pemphigoid antigen 2), is a 180-kDa transmembrane protein within the hemidesmosomal plaque complex, and which is known to be a major antigen in bullous pemphigoid, gestational pemphigoid, cicatricial (mucous membrane) pemphigoid, and linear IgA bullous disease.

Objective

At present, the 3D structure of BP180 is not known. The goal is to predict a reasonable structure for BP180 through machine learning and molecular dynamics.

Methods

In this work, we use the recent Boltz-2 model to predict a putative structure for the intracellular, transmembrane, and proximal extracellular domains, including the NC16A antigenic region and a portion of its first extracellular collagenous domain, Col-15. We computationally embed BP180 in a simple phospholipid bilayer, demonstrate that the putative structure is stable using molecular dynamics, and analyze its allosteric properties.

Results

The structures presented satisfy symmetry and secondary structure properties which are expected from homology modelling. Over three 500 ns trajectories, there is minor instability of the predicted globular head domain, but the homotrimer otherwise stays mostly folded. The putative NC16A domain is stiff, whereas the truncated Col-15 domain is highly flexible. There does not appear to be a nearby stable conformation distinct from the initial state.

Conclusion

The structure presented is a useful starting point for targeting BP180 pharmacologically, for further experimental characterization of BP180, and for generating hypotheses regarding the relevant epitopes contributing to bullous disease. Diffusion models such as Boltz-2 and AlphaFold3 are useful, but their results must be evaluated carefully.

**Word count:** 3131

## Introduction

Human BP180 is a large transmembrane protein 1,497 residues in length, encoded by the COL17A1 gene [1,2]. It is known to exist as a homotrimer at the basal keratinocyte hemidesmosomal plaque, enabling its extracellular domains, which contain multiple collagenous Gly-X-Y repeats, to form the classic collagen triple helix [3]. BP180's first extracellular domain NC16A (residues 490-562) is non-collagenous [4,5], and is known to be the major antigenic region in bullous pemphigoid (BP) in a concentration-dependent fashion [6]. NC16A is also involved in pemphigoid gestationis/gestational pemphigoid [7,8]. Additionally, autoantibodies to BP180 play a role in cicatricial/mucous membrane pemphigoid [9,10] and linear IgA bullous disease [11].

The organization of the hemidesmosome has been known since at least 2003 [12]. Briefly, the intracellular globular portion of BP180 interacts with BP230 (also known as dystonin or BPAG1), to which keratins 5 and 14 are anchored. Additionally, BP180 associates with α6β4 integrin, which itself interacts with plectin, which is anchored to keratins 5 and 14. The hemidesmosome plaque consisting of keratins, plectin, BP180, BP230, and α6β4 integrin thus anchors the intracellular keratin cytoskeleton of the basal epidermis to the epidermal basement membrane (see e.g. Figure 9 of [12]).

Van den Bergh et al. showed that the NC16A region is essential to initiating formation of the extracellular collagen triple helix [4], and therefore, autoantibodies in bullous pemphigoid to the NC16A region interfere mechanically with collagen fibril homotrimer formation of BP180's extracellular domains in an N- to C-terminal direction [13]. In contrast, the classical collagen triple helix is formed in the opposite direction, from the C-terminal to N-terminal [14]. The extracellular portion of BP180 consists of 15 collagenous domains, interspersed by 16 non-collagenous domains [15,16], which potentially provide significant flexibility to the extracellular collagen fibrils.

Because of its length, transmembrane nature, and existence as a homotrimer, BP180 is difficult to study, both biochemically and computationally. Crystallizing BP180 in the absence of a lipid bilayer may yield a biologically irrelevant conformation of the protein, whereas *ab initio* computational folding of the large, membrane-associated homotrimer poses a significant computational barrier.

Recent advances in deep neural networks for multimeric protein folding [17–19] enable putative structures to be generated and further examined using molecular dynamics simulations. In this work, we use such a structure generated by the Boltz-2 model [17] as the starting point for molecular dynamics simulations, investigating the stability and allostery of homotrimeric BP180.

## Materials and Methods

### *Initial structure generation using Boltz-2*

Three copies of the first 600 residues of BP180 (UniProt ID Q9UMD9) were used as input for the Boltz-2 neural network at default settings [17], which is available as a web application from Tamarind Bio (https://tamarind.bio). Folding of this input was also attempted using AlphaFold3 [20], available as a web application from Google DeepMind (https://alphafoldserver.com/), however the resulting structure (Supplemental Figure S1) contained large, unfolded domains and appeared biophysically implausible.

*Visualization of surface electrostatic potential*

The surface electrostatic potential was calculated using APBS [21], which solves the Poisson-Boltzmann equations.

*Embedding of the initial structure using CHARMM-GUI*

The output of Boltz-2 was visually inspected for clashes, and uploaded to CHARMM-GUI [22,23]. CHARMM-GUI Membrane Builder was used to generate the structure embedded in a phospholipid bilayer with dimensions 10 x 10 nm, composed entirely of dimyristoylphosphatidylcholine (DMPC), with water outside of the bilayer and 150 mM NaCl (248 $Na^+$ and 296 $Cl^-$ ions), and with excess Cl ions to neutralize charge imbalances in the protein and lipid bilayer at pH 7.4 [24–28].

The overall simulation box was 10.0 x 10.0 x 32.4 nm, with periodic boundary conditions. The resulting molecular system contained three chains of 600 amino acids, 85,715 molecules of water, 228 molecules of DMPC (99 for the intracellular surface, 129 for the extracellular surface), and the above ions, totaling 310,992 atoms.

*Molecular dynamics simulation*

In triplicate, the resulting system was equilibrated in six stages with default parameters from CHARMM-GUI Membrane Builder, and subsequently simulated for 500 ns at 310 K in GROMACS [29,30], using the CHARMM36m force field [31,32] and TIP3P water [33,34]. The LINCS algorithm was chosen for covalent hydrogen bond constraints. The equations of motion were integrated in 2 fs timesteps for a total of 500 ns of simulated time. Full simulation parameters for equilibration and simulation are included in the Supplemental Data.

*Molecular dynamics trajectory analysis*

Snapshots of the trajectories were taken every 1 ps (500 time steps). GROMACS was used to extract the protein and lipid trajectories, with BP180 centered. The BP180 homotrimer was analyzed in multiple fragments: the globular head (residues 1-489), the NC16A domain (residues 490-566), and the partial Col-15 domain (residues 567-600). The following analyses were performed using the MDAnalysis Python package [35,36].

Each trajectory can be represented as an array of positions $X$ with dimensions of $N_{\text{frames}} \times N_{\text{atoms}} \times 3$. Let $i, j, k$ index these three dimensions. For each protein fragment above, its trajectory was aligned [37] to the first frame containing the $1 \times N_{\text{atoms}} \times 3$ atomic position matrix $X_0$, and the root mean square deviation (RMSD) for a given frame with position matrix $X$ was computed as:

$$\mathrm{RMSD}_i = \sqrt{\frac{1}{N_{\mathrm{atoms}}} \sum_{jk} (X - X_0)_{ijk} (X - X_0)_{ijk}},$$

where the subtraction of $X_0$ is broadcast across the first dimension. The alignments were performed for each: (1) the entire backbone, (2) the globular head (residues 1-489), (3) the NC16A domain (residues 490-566), and (4) the partial Col-15 domain (residues 567-600).

Each trajectory was then aligned (residues 1-600) to its first frame, and root mean-square fluctuations were calculated for each of the α-carbons in the protein backbone using the aligned position array $X$:

$$\mathrm{RMSF}_j = \sqrt{\frac{1}{N_{\mathrm{frames}}} \sum_{ik} (X - \bar{X})_{ijk} (X - \bar{X})_{ijk}},$$

where the angle brackets indicate an average over time, $x_i(t)$ is the vector position of the α-carbon of the $i^{th}$ residue at time $t$, $\bar{X}$ is the time averaged position matrix (reduced over the array's first dimension), and the vertical bars represent the Frobenius norm.

The position matrix $X$ was considered with partially flattened shape $N_{\mathrm{frames}} \times 3N_{\mathrm{atoms}}$, and its time average $\bar{X}$ with shape $1 \times 3N_{\mathrm{atoms}}$, to compute its covariance matrix:

$$C_{mn} = \frac{1}{N_{\mathrm{frames}} - 1} \sum_{i} (X - \bar{X})_{im} (X - \bar{X})_{in},$$

where subtraction is broadcast across the first dimension, and $C_{mn}$ represents the correlation of coordinate $m$ with coordinate $n$. For the covariance analysis, we focused only on the protein backbone, i.e. the nitrogen, α-carbon, and carbonyl carbon. To assign a single covariance per atom, we took a trace over the spatial dimensions. The dimensions of $C_{mn}$ may be reshaped into a tensor of dimensions $N_{\mathrm{atoms}} \times 3 \times N_{\mathrm{atoms}} \times 3$, $D_{ijkl}$., and the traced covariance matrix was defined as:

$$D_{ik} = \sum_{j} D_{ijkj}.$$

Because $D$ scales as the square of the deviation, and we are equally interested in strong negative correlations as positive correlations, we chose to display $\sqrt{|D_{ik}|}$.

For principal component analysis (PCA), the full $C_{mn}$ was diagonalized using PyTorch [38], with the largest eigenvalues $\lambda_m$ and associated unit eigenvectors $\hat{e}_m$ taken as the principal components. The trajectory was projected along $\hat{e}_m$ for visualization:

$$\mathrm{proj}_{\hat{e}_m}(X_{nk}) = (X_{nk} \cdot \hat{e}_m)\hat{e}_m,$$

where the atom position matrix $X_{nk}$ for a given frame is interpreted as a flattened column vector.

Normalized cumulative variances $v_i$ were calculated:

$$v(i) = C \sum_{m=1}^{i} \lambda_m \ ,$$

$$C = \sum_{m=1}^{N_{\mathrm{frames}}} \lambda_m \,,$$

to indicate the fraction of the linear variation in the trajectory that the first $i$ principal components account for.

The three trajectories were concatenated, aligned, and clustered by pairwise aligned RMSD, using gmx cluster at cutoffs of 1.0, 1.5, and 2.0 Å [39].

*Computational resources*

Each simulation was performed on an H100-SXM5-80GB (GROMACS 2020.4) or L48S-40GB (GROMACS 2026.1) NVIDIA GPU, provided by the Bridges-2 cluster at the Pittsburgh Supercomputing Center through the ACCESS program [40,41]. Performance was estimated at 400 timesteps (800 fs simulated time) per second of real time for the H100 GPU, and 700 timesteps (1400 fs simulated time) per second of real time for the L48S GPU. Analyses were performed on a 2020 MacBook M1 Max with 64 GB RAM.

*Visualization and numerical algorithms*

Images and movies of structures were generated using open source PyMol 3.1 [42]. Figures were created with seaborn [43] and matplotlib [44]. NumPy [45], PyTorch [38], and polars (https://pola.rs/) were used to implement numerical algorithms and organize data.

*Code and data availability*

All simulation inputs, parameters, structures, trajectories, and a Jupyter notebook to reproduce figures are included in the Supplemental Code and Data on Zenodo (doi: 10.5281/zenodo.20091510). Instructions are included in a README file in the dataset.

**Results**

The initial machine learning-generated structure by Boltz-2 was visually inspected to confirm (1) $C_3$ homotrimer symmetry, (2) a C-terminal collagen triple helix Col-15, (3) a coiled-coil NC16A domain as first predicted by Balding et al. [3], and (4) an N-terminal globular intracellular domain (Figure 1). In contrast, the AlphaFold3-predicted structure contained large, unfolded domains, even when modeling as a homotrimer (Supplemental Figure S1).

In the AlphaFold Protein Structure Database, the full-length 1,497 residue BP180 monomer (ID: AF-Q9UMD9-F1-v6) is predicted to be a loose collection of α-helices surrounded by low quality unfolded chains (Supplemental Figure S1-A). It is certainly possible that prior to trimerization, BP180 monomers exist in an intrinsically disordered, partially-folded state.

When restricted to the truncated BP180 monomer with residues 1-600, AlphaFold3 predicts a similar unfolded structure. For the BP180 monomer, Boltz-2 predicts a well-folded globular protein (Supplemental Figure S1-C), though now the NC16A and Col-15 domains are enfolded into the globular head. Again, BP180 may exist in such a compact shape after translation and prior to trimerization, however, such a structure is unlikely to be relevant in BP180's biological role, since NC16A and Col-15 are known to be extracellular rather than closely adhering to the intracellular globular domain. AlphaFold3 displays moderate improvement when modeling the homotrimer, now predicting scattered beta-sheets and a collagen triple helix for Col-15; however, the overall structure is still largely unfolded (Supplemental Figure S1-D).

When used to generate structures for just the BP180 ectodomain (residues 490-600), including the NC16A and a portion of the Col-15 domains, Boltz-2 (Supplemental Figure S1-E) and AlphaFold3 (Supplemental Figure S1-F) agreed on a coiled-coil structure for NC16A, which loops back on itself.

Upon visual inspection of the trajectories (Supplemental Movies 1-3), there was mild unfolding of the intracellular domain in trajectory 1. The other two trajectories did not show the same unfolding. To numerically evaluate for simulation convergence and large-scale conformational changes, the root mean square deviations (RMSD), root mean square fluctuations (RMSF), Cartesian coordinate covariance, and principal component analysis of the trajectories were calculated.

The three trajectories demonstrated overall deviation from the initial post-equilibration structure (Figure 2-A), though the absolute change in aligned RMSD was relatively minor, starting near 27 Å and peaking at approximately 31 Å. The deviation appeared to plateau for the latter half of the three simulations. The various domains of BP180 were considered in isolation, with alignments and RMSDs computed independently for each domain. The globular head domain appeared to increase in RMSD (Figure 2-B), whereas the NC16A and Col-15 domains were stable in RMSD (Figure 2-C, Figure 2-D). However, the absolute magnitude for the RMSD of the Col-15 domain was nearly double that of the other domains, stable at approximately 46 Å. The high discordance of the Col-15 domain, even after alignment to itself in isolation of the other domains, is the likely major contributor to the deviation in the overall structure seen in Figure 2-A.

The RMSFs for the three trajectories were consistent with the visual inspection of the trajectories, demonstrating relatively stiff globular head and NC16A domains with RMSFs on the order of 5-10 Å, compared to the highly flexible Col-15 domain (Figure 3), with RMSFs at the C-terminal exceeding 35 Å for trajectory 1. We also remark that the RMSF plot for trajectory 1 is strictly greater than that of the other two trajectories for most residues in the globular head, consistent with mild unfolding of that domain.

Correlations in motion between various parts of the protein were quantified using covariance matrices traced over spatial dimensions (Figure 4), demonstrating strong correlations in motion within and between the three chains of BP180. The eigendecomposition of the untraced covariance matrices yielded the principal components of motion, which are included as .xtc trajectories with accompanying base topology (base.pdb) in the Supplemental Data. Of note, the first principal component of the entire protein backbone appears to involve a bending motion at the NC16A and Col-15 domains (pca_Backbone_0.xtc). The remaining principal components did not contain distinctive features.

Upon clustering with GROMACS at various RMSD similarity cutoffs, we found that 1.0 and cutoffs resulted in 14,938 clusters across 15,001 total structures, whereas the 1.5 and 2.0 Å cutoffs resulted in a single cluster. We report the centroid of the single cluster at 1.5 Å (Figure 5) as the putative structure for the BP180 residues 1-600 homotrimer (clusters_15.pdb in the Supplemental Data).

## Discussion

We first discuss the initial structures which were proposed. Since the extracellular collagenous alternate irregularly with non-collagenous domains, we expected *a priori* that BP180 would form a $C_3$ symmetric homotrimer, which Boltz-2 predicted. Additionally, it has long been predicted from sequence homology that the NC16A domain is a coiled-coil of α-helices [3]. Even in the absence of biochemical evidence of homotrimer formation [15], the characteristic Gly-X-Y collagen repeats of BP180 strongly suggest that BP180 should be considered as a trimer. In this work, the biophysically-informed prediction that BP180 forms homotrimers was essential to generating and differentiating the feasible initial structure generated by Boltz-2 from the unfolded AlphaFold3 prediction.

Addressing now the simulations, the putative BP180 homotrimer structure demonstrated stability across two of three 500 ns, membrane-bound simulations in approximately physiologic conditions, with mild unfolding of the globular head domain in trajectory 1 (Supplemental Movie 1). Since the remainder of the protein remained folded in trajectory 1, and the homotrimer did not dissociate, we interpret the results of these trajectories as indicating the structure proposed is plausible. The homotrimer did not appear to have multiple stable conformations upon clustering.

The globular head domain (intracellular and transmembrane) appears mechanically stiff as does the NC16A domain's coiled-coil of α-helices (Figure 2, Figure 3, Figure 4). As a mediator of cytoskeletal-extracellular matrix mechanical stress, the stiffness of these non-collagenous domains easily communicates mechanical stress across the plasma membrane. In contrast, the partial Col-15 domain was highly flexible, with the first principal component of motion of the homotrimer backbone demonstrating lateral deflection of the Col-15 domain.

Focusing attention closely on the BP180 ectodomain (residues 490-600), which encompasses NC16A (residues 490-566) and a portion of Col-15 (residues 567-600), the Col-15 region appears as a collagen triple helix as expected, both in the initial prediction and throughout the simulations. The NC16A region appears to form α-helices initially (residues 490-505), but with a significant kink at I505 and residues 506-517 form another α-helix. Together, these two α-helix segments with a kink contribute to the predicted coiled-coil appearance of NC16A. The coiled-coil domain is followed by an unfolded domain (residues 518-544), and subsequently a final α-helix (residues 545-566). The various domains within NC16A are interlocked, and remain so over the course of the simulations.

We will now comment on the possible clinical relevance of the proposed structure. It is likely that at least some antibodies in autoimmune bullous pemphigoid are reactive to conformational epitopes in the NC16A domain, which may possess the distinctive structure proposed here. Additionally, there may be linear and conformational epitopes which are hidden due to the high mechanical stiffness and stability of the NC16A domain in this proposed structure. Future biochemical experiments may distinguish the two possible antibody binding mechanisms by comparing reactivity in native versus denatured BP180, which can strengthen the hypothesis that the NC16A domain forms a stiff structure such as that proposed here.

Existing treatments for autoimmune bullous dermatoses rely on immune suppression via corticosteroids or other immune modulators to reduce autoantibodies to cutaneous proteins [46,47]. Furthermore, such therapies are of limited use in non-immune bullous dermatosis resulting from intrinsic genetic defects, such as in a rare case of junctional epidermolysis bullosa [2].

Knowledge of the structures of essential structural proteins in the epidermis may open new avenues of drug development which mechanically reinforce cell-cell adhesion. In the case of bullous pemphigoid due to BP180 NC16A autoantibodies, one could target the NC16A region with a small molecule or peptide to competitively inhibit the autoantibody binding site. Alternatively, one could target the transmembrane or intracellular portions of BP180, with the aim of stabilizing the trimeric state and thereby encouraging proper folding of the collagen triple helix.

For example, tafamidis is an FDA-approved small molecule binder of transthyretin, a tetrameric protein which carries active thyroid hormone [48]. Tafamidis stabilizes the tetrameric structure of transthyretin, thereby preventing its denaturation and subsequent deposition throughout the body in transthyretin amyloidosis. There is also significant interest in manipulating multimeric forms of protein (e.g. lecanemab for amyloid-beta, simufilam for filamin A) to prevent neurodegenerative disease, though with mixed results [49,50].

We will also comment on the use of machine learning in this work. Boltz-2 and AlphaFold3 were both used to generate predictions, however, knowledge of the underlying biophysics and bespoke choice of parameters for molecular dynamics helped to distinguish biophysically-relevant structures from irrelevant ones. Boltz-2 [17] and AlphaFold3 [20] are diffusion/flow-matching models [51,52]. Schematically, given the input sequence $x$, the models may be represented as a neural network $N$ dependent on $x$ and an integration parameter $t \in [0,1]$ such that if the final coordinates of each atom are represented by $F(x,1)$ and some initial coordinates $F(x,0)$,

$$\frac{dF(x,t)}{dt} = N(x,t).$$

The above ordinary differential equation (ODE) may thereby be integrated using numerical integrators such as the first-order Euler method [53]. To generate a variety of conformations for each input, the ODE is upgraded to a stochastic differential equation (SDE) by using the first-order Euler-Maruyama method instead of the Euler method, which adds Gaussian noise during integration [54]. While this approach is mathematically rigorous, there is no intrinsic physical meaning to the parameter $t$, and the architectures of Boltz-2 and AlphaFold3 do not explicitly encode constraints such as "the Col-15 region should be a collagen triple helix," instead relying on training of $N$ or precomputed features such as sequence alignments to appropriately recognize collagenous domains.

Ultimately, this work is computational and requires experimental validation. With that caveat in mind, the BP180 homotrimer structure presented here appears stable, and the computational structures presented possess key features that one would expect from symmetry and homology. In terms of limitations, we modeled only the first 600 residues of BP180, which includes its intracellular, transmembrane, NC16A, and a portion of its Col-15 domains. We did not simulate the remaining low-complexity, extracellular collagenous and non-collagenous domains. We did not model the many other known interacting protein partners, including with the remainder of the hemidesmosomal plaque and with the basement membrane. We also did not model the many post-translational modifications which may occur over time in long-standing structural proteins. To thoroughly explore potential conformation changes, one could perform extensive replica exchange simulations, or observe folding and unfolding events over potentially millisecond-long simulations such as with the Anton-3 supercomputer. Finally, future simulations may also incorporate larger surface area plasma membranes with more accurate lipid compositions.

**Acknowledgments**

Portions of this work were performed while the author was affiliated with Penn State College of Medicine's Medical Scientist Training Program. This work used Bridges-2 at Pittsburgh Supercomputing Center through allocation CIS250721 from the Advanced Cyberinfrastructure Coordination Ecosystem: Services & Support (ACCESS) program, which is supported by U.S. National Science Foundation grants #2138259, #2138286, #2138307, #2137603, and #2138296.

**Declaration of Generative AI Use**

The Boltz-2 generative model was used to produce the initial structure of BP180 homotrimer for simulation. Generative AI was not used elsewhere in this work.

**Figures**

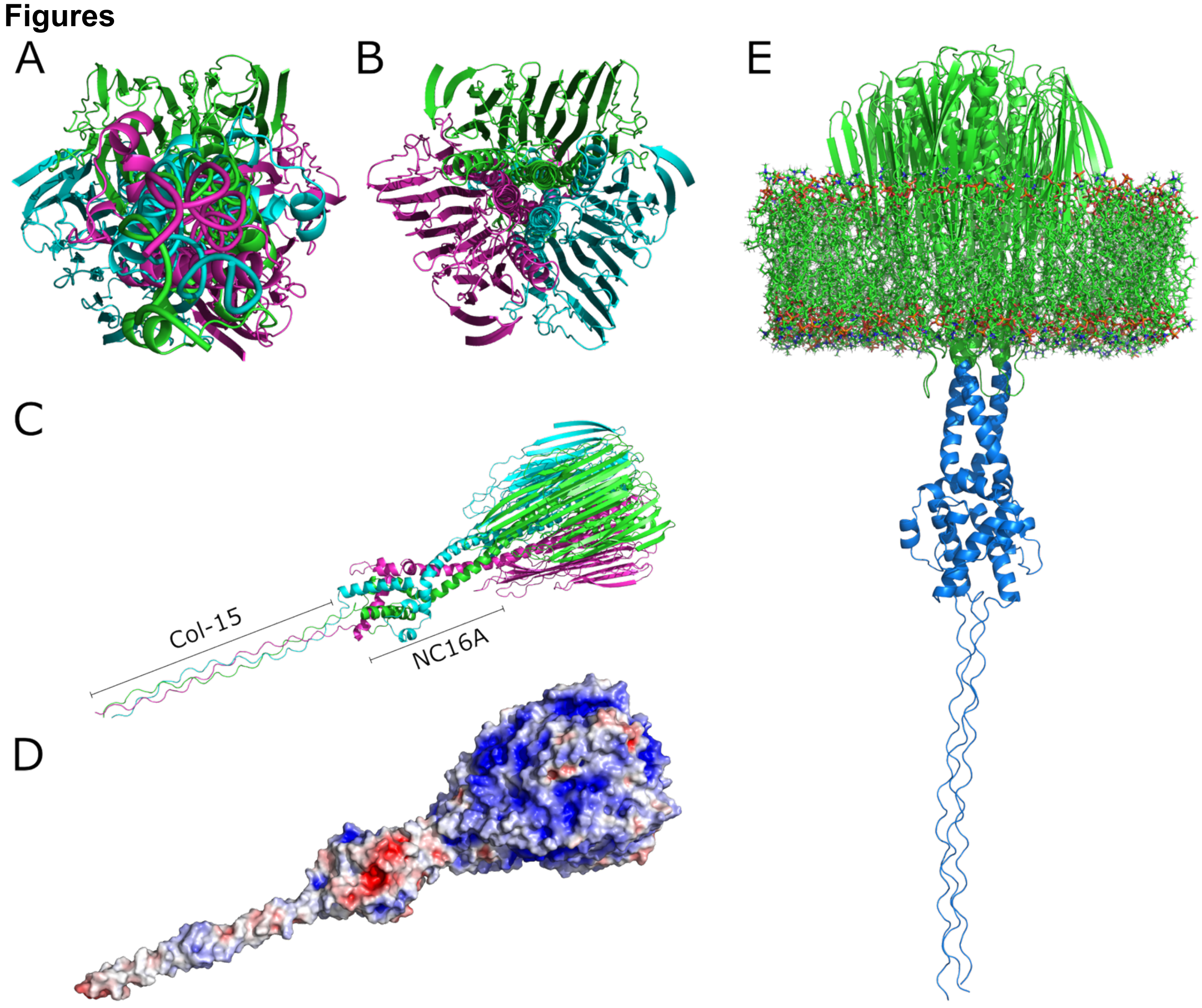

Figure 1: Initial structure of the first 600 residues of BP180 forming a homotrimer, generated by Boltz-2. (A) Bottom-up view (from the extracellular collagenous domain); (B) top-down view (from the intracellular domain); (C) profile view, from the bottom left collagenous domain to the top right intracellular domain; (D) visualization of the surface electrostatic potential, where blue indicates negative potential (-5 $k_BT/e$), red indicates positive potential (+5 $k_BT/e$), and grey indicates neutral potential. Boltz-2 generated a homotrimer with threefold rotational planar symmetry, an extracellular Col-15 collagen triple helix, a coiled-coil NC16A region, and a globular transmembrane/intracellular region. (E) Initial lipid bilayer preparation, generated by CHARM-GUI. Blue indicates residues 490-600, i.e. residues belonging to the NC16A or Col-15 domain. The remainder of the box was filled with rigid water and ions (not shown).

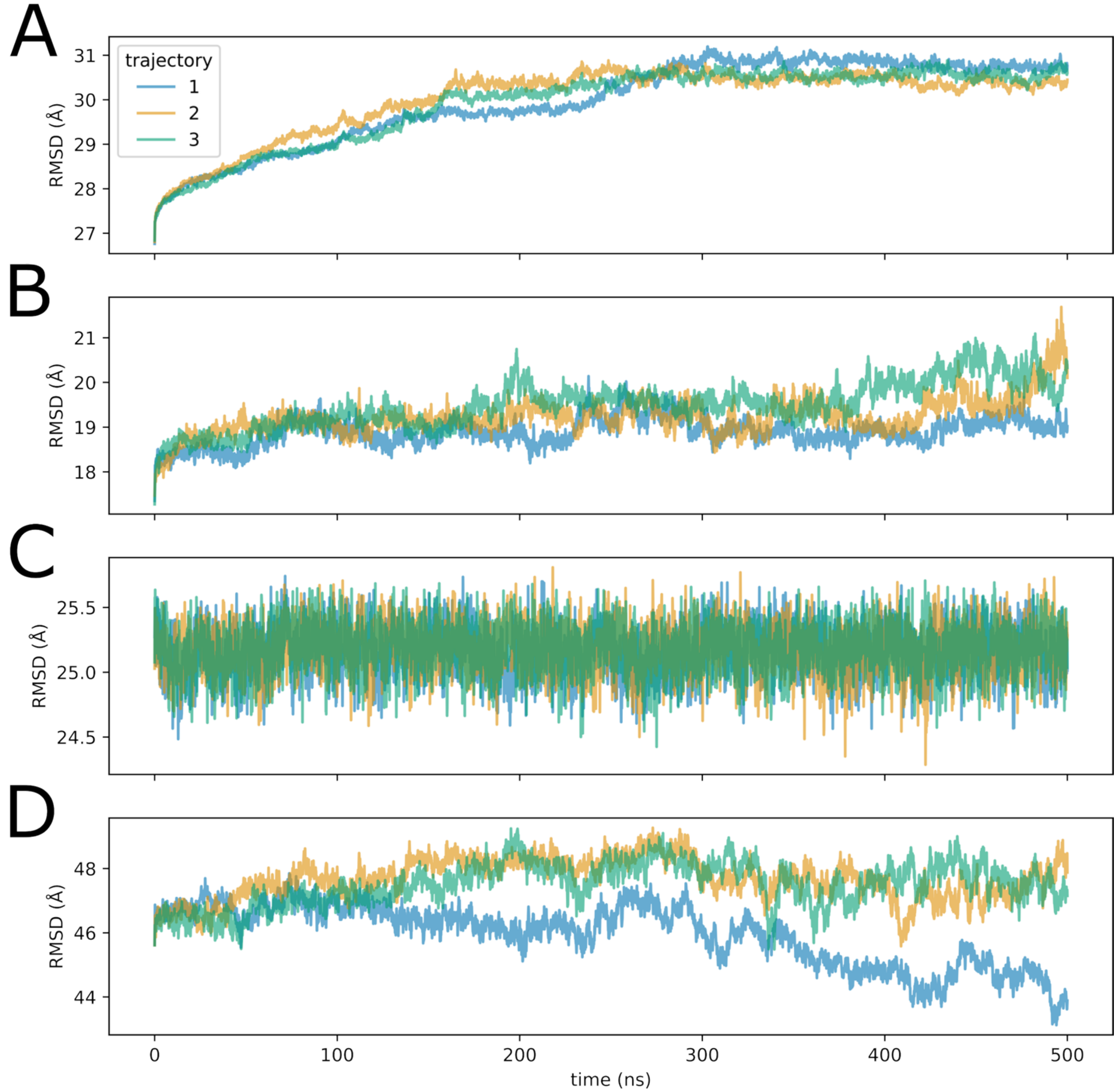

Figure 2: Root mean square deviation over time for the three trajectories, aligned to the post-minimization structure, for (A) the entire homotrimer, residues 1-600; (B) the globular head domain, residues 1-489; (C) the extracellular NC16A domain, residues 490-566; and (D) the partial extracellular Col-15 domain, residues 567-600. RMSDs were calculated with respect to the first frame of the trajectory, following equilibration. The RMSD of all three trajectories appears to increase in the first half of the simulation, but this is due primarily to the highly flexible Col-15 domain, which has an RMSD throughout the simulation of approximately 46 Å.

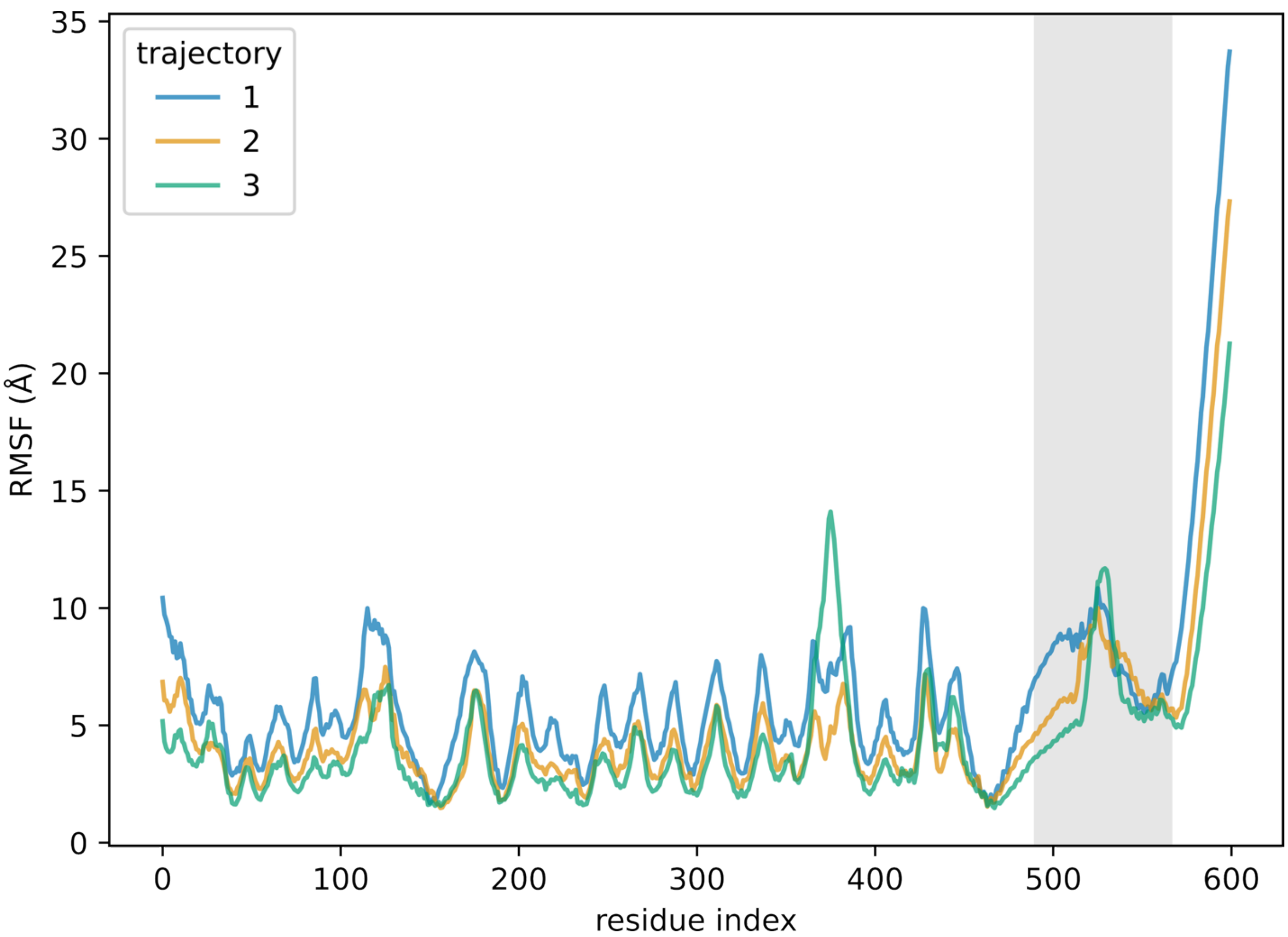

Figure 3: Root mean square fluctuations for each trajectory by residue index, averaged over all three chains of the homotrimer. The relative flexibilities of each residue is similar across all three trajectories. Residues 1-489 corresponding to the globular head domain are relatively stiff, with RMSFs of approximately 5-10 Å. The NC16A domain (residues 490-566, shaded in grey) also appears stiff. The Col-15 domain (residues 567-600) is highly flexible, as expected.

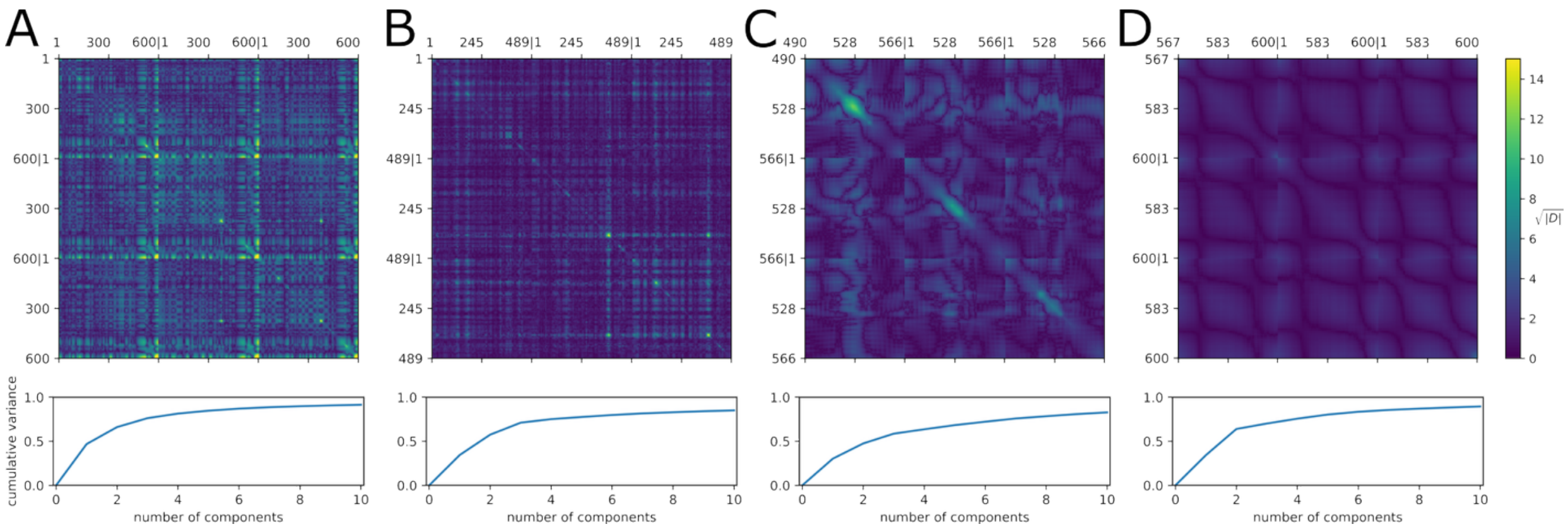

Figure 4: Traced covariance matrices all three trajectories, for (A) the entire protein backbone, (B) the globular head domain, (C) the NC16A domain, and (D) the Col-15 domain, with the strength of correlation ($\sqrt{|D|}$). The axes of the traced covariance matrices are labeled with the residue number within each chain of BP180. The corresponding normalized cumulative variances for the first ten principal components are shown in the bottom row. The matrices are organized into $3 \times 3$ blocks which is visually apparent. As expected, there are strong correlations in motion within each chain (along the block diagonal) and between chains (off-diagonal blocks), indicating high overall stability of the protein.

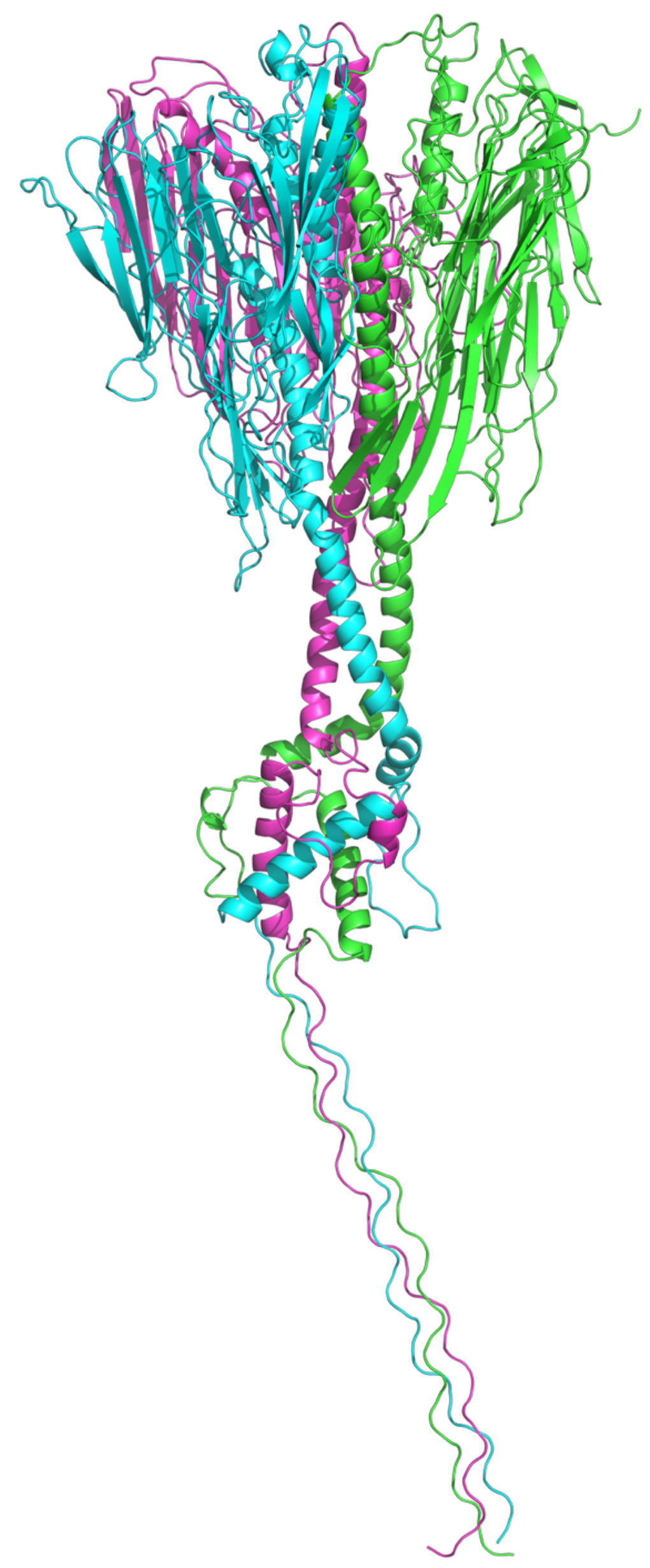

Figure 5: Putative equilibrated structure of BP180 homotrimer.